\documentstyle[12pt]{article}
 
\begin{document}
 
\baselineskip 14pt
 
\begin{center}
{\bf {\large Resonant steps and spatiotemporal dynamics in the damped
dc-driven Frenkel-Kontorova chain}}
 
\bigskip
 
Zhigang Zheng$^{a,c}$%
\footnote{Electronic address: zgzheng@public2.bta.net.cn.} Bambi
Hu$^{a,d}$
and Gang Hu$^{b,c}$\ 
 
{\it $^a$Department of Physics and Center for Nonlinear Studies, Hong Kong
Baptist University, Hong Kong, China}
 
{\it $^b$Center of Theoretical Physics, Chinese Center of Advanced Science
and Technology (World Laboratory), Beijing 8730, China}
 
{\it $^c$Department of Physics, Beijing Normal University, Beijing 100875,
China}
 
{\it $^d$Department of Physics, University of Houston, Houston TX77204,
USA}
 
{\it {}}\bigskip\ 
 
{\bf Abstract}\ 
\end{center}
 
Kink dynamics of the damped Frenkel-Kontorova (discrete sine-Gordon) chain
driven by a constant external force are investigated. Resonant steplike
transitions of the average velocity occur due to the competitions between
the moving kinks and their radiated phasonlike modes. A mean-field
consideration is introduced to give a precise prediction of the resonant
steps. Slip-stick motion and spatiotemporal dynamics on those resonant
steps
are discussed. Our results can be applied to studies of the fluxon
dynamics
of 1D Josephson-junction arrays and ladders, dislocations, tribology and
other fields.
 
\bigskip\ 
 
 
\newpage\ 
 
\section{Introduction}
 
Much attention has been paid in the past twenty years to simple many-body
systems with an effort to disentangle the complexity of macroscopic
systems
with competing interactions. The well-known Frenkel-Kontorova (FK) model,
which describes a chain of atoms interacting with nearest-neighboring
forces
and subject to a periodic substrate potential, is one of the simplest
capable of capturing the essential complexities [1,2,3]. In dimensionless
form, the Hamiltonian of the FK chain reads: 
\begin{equation}
\label{1}H=\sum_{j=1}^N[\frac 12p_i^2+V(x_j)+U(x_{j+1}-x_j)], 
\end{equation}
where $x_j$ denotes the position of the {\it i}-th element in the chain,
and 
$p_i$ is the corresponding momentum. The first term represents the kinetic
energy per element. $V(x)$ describes the substrate potential, which is
assumed to be a periodic form, i.e., $V(x)=V(x+b)$ with $b$ the substrate
period. $U(x_{j+1}-x_j)$ describes the interaction between the nearest
neighbor elements, which is either convex or non-convex, depending on the
studied physical systems. The formula in the Hamiltonian (1) contains both
substrate interactions and mutual couplings between elements, which may
lead
to complicated spatially-modulated structures. Spatially modulated
patterns
have been experimentally observed in many condensed matter physical
systems,
such as ferromagnetic phases of the rare earths and their compounds,
long-period structures of binary alloys, graphite intercalation compounds
or
the polytypic phases of spinelloids, micas, perovskites and other
materials
[4]. In general, the physical origin of this spatially-modulated behavior
is
the competing interactions in the free energy of systems. The FK model is
one of the simplest options among many models which describe this kind of
competitions. The standard FK chain can be described in terms of the
following choices: 
$$
V(x_j)=d[1-\cos (\frac{2\pi x_j}b)], 
$$
\begin{equation}
\label{2}U(x_{j+1},x_j)=\frac 12K(x_{j+1}-x_j-a)^2, 
\end{equation}
where $d$, $K$, $a$ and $b$ denote the potential barrier height, coupling
strength, spring constant and the substrate period, respectively. The
commensurability of the frustration $\delta =\frac ba$ may strongly affect
the spatial structure of the system. The ground state of the FK model was
fully investigated during the last few years, and the
commensurate-incommensurate phase transitions were found and theoretically
described [3]. The theory developed by Aubry [2] stands as one of the
deepest achievements in theoretical comprehension of the physics of
modulated phases. The responses of this system to a dc [5] or an ac or
both
forces [6] in dissipative (inertialess) cases were also explored, where
the
dynamical Aubry's phase transitions and Shapiro steps (dynamical mode
locking) were observed. All these studies reflect the intrinsic properties
of the FK model and reveal some essential features of spatially-modulated
systems.
 
The dynamical FK model was applied to many fields, such as charge-density
waves (CDW) [7], tribology and surface problems [8], self-organized
criticality (SOC) [9] and Josephson-junction arrays (JJA) [10] and
ladders(JJL) [11]. Actually, the FK model is a discretized version of the
sine-Gordon (SG) systems. In an experimental environment, one has to
consider the effects of many external influences, such as dissipations,
fluctuations and external fields, thus it is more reasonable to include
all
these effects. In this paper, we discuss only the spatiotemporal dynamics
of
a damped (inertia) standard FK chain influenced by a dc external force, it
is shown[22] that weak noises can only slightly smooth the resonances
discussed below, and the influence of an ac force will induce more
complicated spatiotemporal patterns. Dynamics in the dissipative
(overdamped) limit under the influence of both dc and ac forces were fully
discussed in relating to CDW problems and the ac effects in experiments of
Josephson-junction ladder arrays [7,11], which were found to correspond to
the dynamical Aubry's phase transition and Shapiro steps, respectively.
However, the inertial effects in some cases might be significant and
cannot
be ignored, moreover, in the underdamped region, inertial effects become
dominant, multistability leads to the so-called hysteresis effects, hence
the system will exhibit complicated spatiotemporal patterns. We will focus
on the underdamped case. The equation of motion discussed here can be
written as
 
\begin{equation}
\label{3}\stackrel{..}{x}_j+\gamma \stackrel{.}{x}_j+\sin
x_j=K(x_{j+1}-2x_j+x_{j-1})+F, 
\end{equation}
where $\gamma $ is the damping coefficient, $K$ the coupling constant and
$F$
the external bias. The frustration $\delta $ does not appear in (3), but
it
plays a significant role in the dynamics of (3). A mechanical realization
is
a chain of $N$ identical damped pendula that are driven by a uniform
torque
and coupled by torsional springs [12]. For very large coupling and number
of
elements, the system (3) can be well described by the continuum SG
equation.
It is shown that [13] when the external applied force varies, the velocity
of the SG chain has a critical value $v_c=2\pi \delta \sqrt{K}$ that
separates two kinds of dynamics (kinks). When $v<v_c$, the motion is that
of
localized solitons, which is called the {\em low-velocity regime}, the
$v-F$
relation in this regime is a continuous line; When $v>v_c$, the motion is
characterized by the whirling wave, i.e., the moving kink is strongly
extended, we call this regime the {\em high-velocity regime}. There exists
an unstable region between these two regimes, which corresponds to the gap
on the $v-F$ characteristics. These two regimes exist for both discrete
cases and continuum cases, but the dynamics introduced by discreteness may
be a distinct feature, which will not happen for the continuum cases. The
whirling-instability induced resonances in the high-velocity regime were
well described in [14]. We will focus on the dynamics in the low-velocity
regime, which is the consequence of another kind of mechanism. We shall
give
a precise mean-field description of resonances in this regime.
 
The paper is arranged as follows. In Sec.II, the dynamics in the
low-velocity regime are theoretically discussed, where we will introduce a
mean-field treatment which is proved to be perfectly effective [15]. This
treatment results in a complete description of the resonance behavior in
the
low-velocity regime. Sec.III contributes to numerical simulations. We show
that the theory proposed in Sec.II agrees very well with numerical results
by varying the mean-field parameter. In Sec.IV, the physical meaning of
the
mean-field consideration is discussed, which is related to Aubry's CI
phase
transition. Sec.V gives a discussion on prohibited resonances and gives a
resonance prohibition criterion. The high-velocity whirling mode will lead
to the solitary-wave instability of some low-velocity steps.
Spatiotemporal
dynamics on low-velocity steps are discussed in Sec.VI. It is shown that
three kinds of motions, i.e., periodic, quasiperiodic and chaotic motions
can be observed on steps. We summarize the results and propose future
topics
in Sec.VII.
 
Numerical simulations will be used to study the spatiotemporal dynamics in
this paper. The fourth-order Runge-Kutta integration algorithm is used and
the time step is adjusted according to the numerical accuracy. Periodic
boundary conditions are added, i.e., $x_{j+N}(t)=x_j(t)+2\pi M,$ where $M$
is an integer that counts the net number of kinks trapped in the ring,
therefore the frustration is $\delta =\frac MN$ and the equilibrium
distance
between the elements in the absence of the substrate will be $a=2\pi
\delta $%
. We mainly discuss the average velocity of the chain, which is a good
candidate for studying the response of the FK chain to external forces.
The
averaged velocity of the chain is defined as $v=\frac 1N\sum_{j=1}^N<%
\stackrel{.}{x}_j>$, where $<.>$ denotes the time average.
 
\section{Kink-radiation induced phase locking: resonant steps}
 
The key consequence introduced by discreteness of the chain is that the
solitary wave will radiate small-amplitude linear waves when it moves. The
mechanism behind this behavior is the competition between the harmonic
chain
and the periodic substrate. Due to the discreteness of the chain, it will
collide with the substrate when it moves. For the continuum SG systems,
the
attractors in the low-velocity regime are travelling waves. Such is also
the
case for the discrete version, while the wave is composed of a moving kink
and its radiated phonon waves in its wake. This is shown in our numerical
experiments. The phenomenon of the radiation by a moving kink was
discussed
by Currie {\it et al}.[16], Peyrard and Kruskal [17] and other authors
[18]
in numerical studies, and also found in experiments on Josephson-junction
arrays[10]. In some cases the kink motion and its radiated waves can
become
phase-locked and then lead to quantized velocity of the chain under a
constant force. This occurs when the linear modes are separately excited,
for if many different modes are excited simultaneously the resonance will
overlap and then discrete velocity cannot be observed. It should be noted
that the kink shape will strongly affect the final results, as can be seen
below.
 
Theoretical considerations were explored by several authors [10,14], but
some drawbacks and even mistakes exist in their discussions. {\bf First},
the equation of motion was directly linearized to discuss the linear waves
radiated by moving kinks. One should be aware that the linearization
should
be used around the {\em moving kink}, thus the direct linearization is not
correct. {\bf Second}, an approximation of the kink directly by a $2\pi $%
-form, this consideration is too crude to grasp the crucial points. In
fact,
only in some limiting cases, for example, when $M=1$ and $N$ is very
large,
i.e., $\delta \rightarrow 0$, this approximation is valid; For finite
frustration $\delta $ , the kink apparently is not a $2\pi $ form, thus
the
effect of the kink solution should be taken into account. {\bf Third}, all
previous discussions did not consider {\em sub-harmonic resonances}, which
may be very important for finite frustration (multiple trapped kinks)
cases,
thus the resonance condition should include the subharmonic cases. Under
the
considerations of all above points, let us give a more precise description
of quantized velocities.
 
Assuming the static kink solution is $\left\{ x_j^{*}\right\} $,
$j=1,...,N$
, when it moves along the chain, linear phonon waves is radiated in its
wake
due to the discreteness, then we can linearize the equation of motion (3)
around the moving kink $x_j^{*}(t)$ by inserting
$x_j(t)=x_j^{*}(t)+u_j(t)$
into (3) and get 
\begin{equation}
\label{4}\stackrel{..}{u}_j+\gamma \stackrel{.}{u}_j+[\cos
(x_j^{*})]u_j=K(u_{j+1}-2u_j+u_{j-1}).
\end{equation}
For sufficiently small damping $\gamma $, we can neglect the dissipative
term in (4) and consider the conservative case. In fact, the dissipation
effect can be compensated by the external driving force. We intend to get
the dispersion relation of the linear phonons. The difficulty in (4) lies
in
the {\em lattice dependence of the kink solution} $x_j^{*}$. An
alternative
way to overcome this difficulty is to introduce the {am\ mean-field
treatment%
}, i.e., we replace the term $\cos (x_j^{*})$ by an averaged quantity 
\begin{equation}
\label{5}\beta =\frac 1N\sum_{j=1}^Ncos(x_j^{*}),
\end{equation}
where we call the parameter $\beta $ a {\em contraction factor}. One
should
note that here $\beta $ is time independent, because it depends on the
{\em %
topological structure of the static kink} (We also find the dependence of
$%
\beta $ on the external driving $F$, leading to additional contraction
effect. But the present approximation already gives a very precise
prediction. This effect will be discussed elsewhere.). In fact, this
parameter does describe the contraction effect of the kink solution. For
$%
\delta \rightarrow 0$, e.g., $M=1$ and very large $N$, the kink solution
can
be approximated by a $2\pi $ form: 
\begin{equation}
\label{6}x_j^{*}(t)=\left\{ 
\begin{array}{c}
0,
\mbox{ for }j<vt, \\ 2\pi ,\mbox{ for }j>vt.
\end{array}
\right. 
\end{equation}
Then we immediately get $\beta =1$ by using (5). This is the case
discussed
by Watanabe {\it et al}., but the story does not end here. As will be
shown
below, the frustration plays a crucial role in the chain dynamics. Now let
us insert the linear phonon mode $u_j(t)=\exp [i(\omega _lt+kpj)]$ into
(4)
and use (5) instead of the lattice dependent term $\cos (x_j^{*})$, we may
get the dispersion relation 
\begin{equation}
\label{7}\omega _l=\sqrt{\beta +4K\sin {}^2(\frac 12kp)}.
\end{equation}
The circulation frequency of the moving kink is $\omega _k=\frac{2\pi
}T=<v>$
, the resonance between these two frequencies leads to the discrete
velocity
of the chain. The resonance condition reads 
\begin{equation}
\label{8}m_1\omega _k=m_2\omega _l,
\end{equation}
when the kink rotating frequency and its linear mode become phase locked.
The geometry constraint (periodic boundary condition is applied in this
paper) implies that the wave-number should satisfy $kp=2\pi \delta
m_1/m_2$.
Then we get the resonant velocity steps
 
\begin{equation}
\label{9}v(m_1,m_2)=\omega _k=\frac{m_2}{m_1}\sqrt{\beta +4K\sin
{}^2(\frac{%
m_1\delta \pi }{m_2})}, 
\end{equation}
where $(m_1,m_2)$ is an integer pair that describes the resonance between
kinks and linear waves, $\delta $ is the frustration. The result we
obtained
here considers all the points we mentioned above, which gives a complete
resonant velocity spectrum. The significance of (9) lies in that it
considers the commensurability effect induced by the winding number
$\delta $%
, and moreover this formula relates only to the frustration $\delta $,
implying that the resonance is not a finite-size effect, but the
discreteness effect. The resonance condition (8) is the main point,
however,
it is very strange that all authors only used the integer resonance
$m_2=1$.
In fact, one may frequently observe higher order resonant steps, as shown
below. By considering both the correct resonance criterion and the
commensurability effect, formula (9) gives a complete description of all
possible quantized velocities.
 
\section{Numerical results}
 
We performed numerical simulations of the system (3) for different number
of
particles and different other parameters. In Fig.1, the $v-F$ relations
are
given for $N=8$, $\gamma =0.1$ and different number of trapped kinks
$M=1$,
2 and 3. First we notice that all curves have resonant steps (plateaus),
which are the consequences of the locking between the moving kink and its
linear phonons. In many regions a given driving force corresponds to
several
velocities, indicating the existence of multistability. Hysteresis can be
observed when one adiabatically changes the external driving force, which
is
a direct consequence of many attractors (metastable states). Keeping in
mind
that for the continuum SG system, the $v-F$ curve in the low-velocity
regime
is a continuous line, the transition between resonant steps will result in
the discontinuousness of the velocity line, replaced by many quantized
values. We label the resonances on steps by a pair of integers $(m_1,m_2)$
in terms of the theoretical formula (9), all the steps can be well
predicted
by using (9) if we choose $\beta =$0.55, 0.25 and 0.15 for $M=$1, 2 and 3,
respectively. Subharmonic resonances can be frequently found for $M>1$,
i.e., these resonances can be easily excited and locked when there are
more
than one kink trapped in the ring. To make a clearer comparison, we give a
table of resonances observed in Fig.1. It can be clearly shown that
formula
(9) we obtained above gives not only a much better prediction of $(m_1,1)$
resonances than previous prediction $v=\frac 1{m_1}\sqrt{1+4K\sin
{}^2(m_1\delta \pi )}$ [14] but also precisely predicts high-order
resonances.
 
Fig.2 shows several modes of motion of one of the particles in the chain
for
different cases. In (a), we give an evolution of a 16-particle chain in
one
period with only one kink trapped inside and for a small drive. Only the
positive part is shown. It is observed that the velocity have a higher
peak
and some lower peaks. The higher peak indicates that the particle hops
from
one potential well to another one, and the nearest lower peak is the
influence of the coupling indicating the hopping of the nearest neighbor
particle. Obviously the propagation of the coupling effects decays
exponentially, then the chain exhibits a slip-stick motion. With
increasing $%
F$, the influence of adjacent particles becomes large. (b) gives a 5:1
resonance for $N=8$ and $M=3$, where the particle hops once in each period
that contains five peaks. Additionally, the multi-kink effect becomes
evident, which is shown in (c) for a larger force. This is a higher-order
resonance.
 
One should be aware that each point on the $v-F$ curve perhaps corresponds
to several attractors, not just a single state as pointed by other
authors.
One may find both $m_1:m_2$ and $nm_1:nm_2$ ($n>1$) resonances at the same
value of the driving force when starting from different initial
conditions,
which correspond to the same value of the average velocity. This indicates
that many kinds of spatiotemporal patterns can be found. Moreover, when
the
resonant values are very close to each other, the velocity steps are
almost
degenerate and these very close modes may also interact with each other to
result in complicated motions.
 
For the weaker coupling strength $K$, more resonances can be excited. In
Fig.3, we give the $v-F$ plot for different couplings $K=$2, 1, 0.25 and
0.1. One finds the steps can be easily distinguished from each other for
larger $K.$ When $K$ decreases, resonances will become vague and
overlapped
so that the steps cannot be well distinguished. For $K\rightarrow 0$, the
behavior of the chain approaches that of uncoupled particles, thus
bistable
solutions can be observed [19].
 
An interesting problem relates to the dynamical manifestation of an
incommensurate chain, for example when the winding number $\delta $ is the
golden mean $\delta _G=(\sqrt{5}-1)/2$. We mentioned that the resonant
steps
are only related to the winding number and independent of the number of
particles. One may use the Fibonacci sequence to approach the golden mean,
i.e., $\delta =\frac 58,$ $\frac 8{13},$ $\frac{13}{21},$ $\frac{21}{34},$
...$\rightarrow \delta _G$. In Fig.4, the $v-F$ characteristics is
numerically calculated for the Fibonacci approach to the golden mean.
Different damping parameters are adopted, but this will not affect the
value
of the step. This is proved in numerical computations. One can easily find
the main resonant steps coincide with each other for different $N$ and
$M$.
This result proves that the resonant steps depend only on the winding
number 
$\delta =\frac MN.$
 
\section{Physical interpretation of the mean-field treatment}
 
It is shown from the above discussions that our theoretical result can
precisely predict all possible resonant steps, but one may find that
$\beta $
varies for different frustration $\delta $. Then a natural question
arises:
what is the relation between the contraction factor and the winding
number?
Does the contraction factor have some physical meaning? We focus on the
exploration of this problem.
 
Because the property of $\beta $ depends on the form of the static kink,
let
us study the dissipative case: 
\begin{equation}
\label{10}\stackrel{.}{x}_j=-\cos (x_j)+K(x_{j+1}-2x_j+x_{j-1})+F. 
\end{equation}
In the absence of $F$, (10) is used to numerically explore the ground
states
of the FK chain [2]. In fact, the static kink just corresponds to the
ground
state, which was fully studied in recent years. Bearing this in mind, let
us
recall the studies of the dissipative dc-driven dynamics of the FK chain.
It
was found that [5] there exists a critical depinning force $F_c$, below
which the chain will be pinned and above which the chain will slide. For a
commensurate chain (rational $\delta $), $F_c>0$ for all $K$, while for
the
irrational $\delta $, there exists a critical $K$, when $K>K_c$ the chain
is
pinned and when $K<K_c$ the chain slides. By averaging (10) on the lattice
label $j$ , one can get 
\begin{equation}
\label{11}<\stackrel{.}{x}_j>=-<\cos (x_j)>+F, 
\end{equation}
only when $F>F_c$ can the static kink move freely along the chain, hence
we
readily get 
\begin{equation}
\label{12}\beta =<\cos (x_j^{*})>=F_c, 
\end{equation}
that is, $\beta $ is the depinning force that is needed to overcome the
Pierels-Nabarro (PN) barrier and move continuously the static kink along
the
chain [2]. This connection is very interesting because knowledges of
Aubry's
CI phase transitions now can be directly applied to this dynamical case.
For
example, $\beta $ is directly related to the lowest phonon frequency for
the
FK eigen spectrum by the scaling relation $\beta \propto \omega _G^2$.
 
In Fig. 5, we give the relations between the contraction factor $\beta $
and
the frustration $\delta $ for different coupling constants. The curves are
symmetric about 0.5. This is a consequence of the symmetry of the chain.
It
is shown that the curves are not monotonical, i.e., many peaks can be
observed. These peaks are no other than the rational resonances. The most
significant resonances lie at $0:1$, $1:1$ and $1:2$, and other resonances
can also be observed for moderate $K$. In fact, these resonances build a
Farey tree. This means all resonances can be found by using the Farey
sequence: $\frac{p_n}{q_n},\frac{p_{n+1}}{q_{n+1}}\rightarrow \frac{%
p_n+p_{n+1}}{q_n+q_{n+1}}$. With increasing coupling K, the curve becomes
lower, and the contraction factor for the golden-mean winding number
$\delta 
$ approaches 0. It is expected that there exists a critical $K_c$, when $%
K>K_c$, $\beta =0$ for $\delta _G$. In fact, Aubry's transition from the
pinned state to a sliding state occurs. In this case, the kink has the
translational symmetry, thus the summation can be replaced by the integral
for large $N$ in (5), $\beta =\int_0^{2\pi }cos(x^{*})dx^{*}=0$. In the
previous section, the values of $\beta $ we chose to predict resonant
steps
agree very well with the $\beta -\delta $ line $K=1.0$. This strongly
supports the interpretation of $\beta $. It is interesting that a similar
plot between the depinning force and the frustration was obtained
numerically and experimentally by Ustinov {\it et al.} [10] for
explorations
of the fluxon dynamics in the Josephson-junction arrays, where the
frustration corresponds to the magnetic field. This also confirms our
conclusion.
 
The maximum velocity occurs on the 1:1 resonant step, i.e. 
\begin{equation}
\label{13}v_{\max }(\delta ,K)=\sqrt{\beta (\delta ,K)+4K\sin {}^2(\delta
\pi )}, 
\end{equation}
which is a function of $\delta $ and $K$. Because the contraction factor
$%
\beta $ is a highly nonlinear function of the frustration $\delta $, the
relation between the maximum velocity and the frustration is also
complicated. In Fig. 6 we give the relations $v_{\max }-\delta $ for
different coupling strengths $K.$ It can be found that the relation is
rather complicated for small $K$, where the contraction factor can not be
neglected. While for stronger couplings, the dominant factor is the
sinusoidal term, where the relation is approximately sinusoidal, $v_{\max
}\propto \sin \delta \pi $. But at the edges of all the curves, the
relations still are nonlinear, where $\beta $ plays a dominant role.
 
\section{Resonance prohibition criterion}
 
The agreement between the theoretical and numerical results is quite good,
as found from the above discussions. One may also find that not all the
resonances can be observed from the numerical curves, i.e., the resonances
are incomplete. Several reasons can lead to the disappearance of some
steps. 
{\em First}, the dissipative effect should be taken into account. Large
damping $\gamma $ can smooth those weaker resonances. {\em Second}, the
coupling strength also plays a role in the disappearance of some weaker
resonances, this can be easily understood from the SG dynamics. For
stronger
couplings, the discreteness of the chain becomes weak. However, the above
two effects are all parametric, i.e., whether the steps can be observed
strongly depends on the parameter region one chooses. In fact, a {\em
third}
reason, an intrinsic reason, should not be ignored. That is,
kink-radiation
induced resonances can only occur in the low-velocity regime. Therefore a
criterion should be 
\begin{equation}
\label{14}v(m_1,m_2)\leq v_c=2\pi \delta \sqrt{K}, 
\end{equation}
which means that the anticipated steps that exceeds $v_c$ will not be
observed, where $v_c$ is the boundary between the low- and high-velocity
regimes, which is pointed out in Sec.I. By inserting (9) into (14) and
setting $X=\frac{m_1\delta \pi }{m_2}$, where $X\in [\delta \pi ,\infty )$
(because $m_2\leq m_1$), we may obtain the following form, 
\begin{equation}
\label{15}X^2-\frac \beta {4K}\geq \sin {}^2X\mbox{ and }X\in [\delta \pi
,\infty ). 
\end{equation}
In Fig.7 we plot the two functions $f(x)=X^2-\frac \beta {4K}$ and $\sin
{}^2X$ for $\beta =1$, $0.5$ and 0 (dotted lines from lower to upper),
respectively. For $\beta =0$, the point of intersection of the two curves
is
0, above which the inequality satisfies automatically. In this case, all
possible resonances can be found. $\beta =1$ occurs for $\delta
\rightarrow
0 $ or 1, where the intersection point is approximately $X_0=0.95$. In
numerical simulations, we use usually a finite number of particles, thus
the
frustration $\delta $ is fairly small when $M$ is very small and $N$ is
very
large. For example, in the case $\delta =1/N$, there are some disappeared
main steps($m_2=1$). The disappeared steps satisfy $m_1\delta \pi =\frac{%
m_1\pi }N\leq X_0$, which leads to $m_1\leq NX_0/\pi \simeq 0.95N/\pi $.
For 
$N=8$, $m_1<3$, which is in good agreement with the observation of Fig.1.
In
Fig.8, we give the numerical $v-F$ characteristics for $N=16$, $M=1$. Both
the low- and high-velocity regimes are shown. The steps above the dotted
horizontal line are whirling-mode-instability induced resonances, whose
mechanism is different from resonances found in the low-velocity regime.
Arrows show the jumps between the low- and high-velocity regimes. The
inset
of Fig.8 is an enlarged plot of steps in the low-velocity regime. One may
find that resonances start only from $6:1$. Resonances $m_1:1$ satisfying
$%
m_1<6$ may immerse in the high regime, which become unstable due to the
whirling modes. By using our criterion (15), we get $m_1\leq 5$. This is
the
same as the numerical result in Fig.8.
 
The critical frustration that indicates the disappearance of resonant
steps
can be determined by 
\begin{equation}
\label{16}v_{\max }=v_c=2\pi \delta \sqrt{K}. 
\end{equation}
When $v_{\max }<v_c$, all the steps can be observed. When the frustration
$%
\delta $ is less than a critical $\delta _c,$ $v_{\max }>v_c$, some steps
begin to immerse into the high-velocity regime and cannot be observed. The
critical frustration cannot be theoretically worked out from (16). We can
only obtain it from the numerical computations. For $K=$0.5, 1.0 and 10.0,
we obtain $\delta _c=0.30$, 0.222 and 0.1010, respectively.
 
\section{Dynamics on steps}
 
The attractors on steps are deformed travelling waves, and the
phase-locking
between the kink motion and the linear waves leads to resonant steps. This
basic mechanism is quite simple, but the spatiotemporal dynamics on steps
may be sophisticated. Dynamics on high wave numbers (large $m_1$) is a bit
simple, because they can be easily excited by a small external force. For
larger forces, many linear modes can be excited and they may also interact
with each other, thus the dynamics become complicated. A quite interesting
problem relates to the dynamics of the transitions between resonant steps.
In numerical simulations, we observed three routes of motions along a step
when one adiabatically increases the driving force $F$:
 
{\em Route (1)}: periodic$\rightarrow $periodic$\rightarrow $transition to
new steps;
 
{\em Route (2)}: periodic$\rightarrow $quasiperiodic$\rightarrow
$transition
to new steps;
 
{\em Route (3)}: periodic$\rightarrow $quasiperiodic$\rightarrow
$chaotic$%
\rightarrow $ transition to new steps.
 
Route (1) mainly occurs for small $F$ cases. In this case, on a resonant
step, a moving kink couples to its linear waves and they are phase locked.
Further increase of the driving force will not further increase the
velocity
of the chain, the increase of energy is used to {\em amplify the linear
waves%
}, then at a critical force the large linear wave may cause the resonance
unstable, the transition to another step occurs. Here the coupling to
linear
waves can be considered as an {\em additional damping}. When the external
drive becomes larger, several linear modes will interact with each other,
leading to the quasiperiodic or even chaotic motions. The role of the
drive
in these cases will be both amplifying the linear waves and {\em exciting
new linear modes}. In Fig. 9, we calculated Poincare sections $\sin
x_1-\sin
x_2$ by strobing the phases whenever the mean phase $\stackrel{-}{x}%
(t)=\frac 1N\sum_{j=1}^Nx_j(t)=0$ (mod $2\pi $) for $N=8$, $M=2$ on the
$7:3$
step. From (a) to (e), $F=$0.112, 0.12, 0.13, 0.15, 0.16 and 0.165,
respectively. For a smaller $F$, the chain exhibits periodic motions, the
section only contains a point, this is not shown in Fig.9. For $F>0.11$,
the
motion becomes quasiperiodic, corresponding to the emerge of a small
closed
torus in (a). Further increases of external force cause the quasiperiodic
motion more complicated, the closed curve becomes {\em deformed, twisted
and
enlarged}, indicating that the energy provided by the external drive is
consumed for both exciting new linear phonons and amplifying linear waves.
When $F$ exceeds a critical value, the torus breaks, a new periodic motion
emerges again, then a new resonance occurs, which corresponds to the
transition to a new resonant step. We also show a route to chaos in Fig.10
for $N=8$ and $M=3$ on the $3:2$ step, where $F$ varies from 0.162 to
0.213
for the plots from (a) to (l). This cartoon vividly demonstrates the route
from the periodic to chaotic motions. We still did not plot the periodic
motion, for it is only a point on the section. Then the point grows and
becomes a torus, indicating the presence of quasiperiodic motion. The
torus
becomes larger and twists, like the presence of a ''{\em saddle}'' point
(it
is not really a saddle), then more ''saddles'' occur in (g) and the torus
becomes a {\em web}. Further increasing $F$ leads to the irregular motion.
The motions in (k) and (l) become chaotic. In fact, one may study the
interaction of linear modes on these steps. This problem now is under
further study.
 
\section{Conclusions}
 
This paper deals with the discreteness effect of the SG chain in the
low-velocity regime. Small linear phonons can be radiated when the
localized
kink moves along the chain in this region. When the velocity of the moving
kink and its linear-wave frequency satisfy the resonance condition, the
motion will be phase-locked, then one can observe the resonant steps in
the $%
v-F$ plot. We give a mean-field treatment of the resonances and give a
complete description of resonant steps. Our theoretical formula can
precisely predict all the steps observed in numerical simulations. In
fact,
we show that the mean-field approximation is physically reasonable, the
contraction factor corresponds to the depinning force that is necessary to
overcome the PN barrier and move freely the static kink along the chain.
This directly connects the present results to Aubry's CI phase
transitions.
Commensurability plays a crucial role in the kink dynamics of the discrete
FK chain. Due to the gap between the low- and high-velocity regimes, in
some
cases not all the steps can be observed, some of which are prohibited to
occur because they immerse into the high-velocity regime and becomes
unstable. We derive a step prohibition criterion, which agrees very well
with numerical observations. We also carefully investigated the dynamics
on
steps and found three kinds of dynamics of transitions between the
resonant
steps. Dynamics on low steps are generally simple, only a single linear
mode
will participate in the competition, thus the motion is periodic; On
higher
steps, several linear modes can be simultaneously excited and they compete
and interact with each other, leading to quasiperiodic and even chaotic
motions. Every jump from one step to a higher step can only produce
periodic
motion, and when further increasing the driving force the linear wave is
amplified and other modes are also excited, leading to complicated
spatiotemporal patterns.
 
The damped dynamics of the FK model can be found in many fields of
physics,
for example in CDW, coupled damped pendula, JJL and JJA. Special interests
were focused on experimental studies of fluxon dynamics in JJA and JJL in
recent years. Josephson junction is an excellent candidate for the study
of
nonlinear dynamics and it has been applied in many fields. Resonant steps
were also observed in the experiments of the current-voltage
characteristics
for JJL and JJA. Therefore we expect the present results can also be
applied
to experiments on JJA and JJL. For the JJL, the coupling mechanism is very
complicated, the coupling usually is a sinusoidal form, thus only for
larger
coupling, our results can be applied to experiments of JJL [22].
 
The noise effect on the resonant steps may cause anomalous diffusion of
the
Brownian motions of damped FK chain [20]. This is quite different from the
diffusion of an uncoupled group of Brownian particles in biased periodic
potentials[19,21]. Another problem relates to the ac-driven dynamics of
the
damped FK model. Because of competitions of many time scales, the dynamics
may be more complicated. This problem is under study now.
 
\bigskip
\bigskip
 
One of the authors(Zheng) thanks all the colleagues in Center for
Nonlinear
Studies of Hong Kong Baptist University for many valuable discussions.
This
work is supported in part by the Research Grant Council RGC and the Hong
Kong Baptist University Faculty Research Grant FRG. It is also supported
by
the National Foundation of Natural Science of China.
 
\newpage\ 
 
\begin{center}
{\bf References}
 
\bigskip\ 
\end{center}
 
1. J.Frenkel and T.Kontorova, {\it Phys. Z. Sowjet.} {\bf 13}, 1 (1938).
 
2. S. Aubry, {\it Phys. Rep.} {\bf 103}, 12 (1984); M.Peyrard and S.Aubry, 
{\it J. Phys.} {\bf C 16}, 1593 (1983).
 
3. W.Selke, {\it Phys. Rep.} {\bf 170}, 213 (1988).
 
4. J.M.Yeomans, {\it Solid State Physics} Vol. 41, edited by H.Ehrenreich
and D.Turnbull (New York: Academic).
 
5. S.Coppersmith, {\it Phys. Rev.} {\bf B30}, 410(1984); S.Coppersmith and
D.Fisher, {\it Phys. Rev.} {\bf A38}, 6338 (1988); L.Sneddon and K.Cox,
{\it %
Phys. Rev. Lett.} {\bf 58}, 1903 (1987); L.Sneddon and S.Liu, {\it Phys.
Rev.%
} {\bf B43}, 5798 (1991).
 
6. L. Floria and F. Falo, {\it Phys. Rev. Lett.} {\bf 68}, 2713 (1992);
F.Falo, L.Floria, P.Martinez and J.Mazo, {\it Phys. Rev.} {\bf B48}, 7434
(1993); J. Mazo, F.Falo and L.Floria, Phys. Rev. {\bf B52}, 6451 (1995);
For
recent reviews on dissipative dc and ac dynamics in FK model, see L.Floria
and J.Mazo, Adv. Phys. 45, 505 (1996).
 
7. A.A.Middleton and D.S.Fisher, Phys. Rev. Lett. 66, 92 (1991);
A.A.Middleton, Phys. Rev.\ Lett. 68, 670 (1992); A.A.Middleton, O.Biham,
P.B.Littlewood and P.Sibani, Phys. Rev. Lett. 68, 1586 (1992).
 
8. B.N.J.Persson, Phys. Rev. Lett. 71, 1212 (1993); B.N.J.Persson, J.
Chem.
Phys. 103, 3849 (1995); T.N.Krupenkin and P.L.Taylor, Phys. Rev. B 52,
6400
(1995); E.Granato, M.R.Baldan and S.C.Ying, Los Alamos preprint No.
cond-mat/9510058 (1995); M.Weiss and F.J.Elmer, Phys. Rev. B 53, 7539
(1996); M.Weiss and F.J.Elmer, Los Alamos preprint No. cond-mat/9704110
(1997).
 
9. F.J.Elmer, Phys. Rev. E 50, 4470 (1994); F.J.Elmer, Helv. Phys. Acta
66,
99 (1993).
 
10. H.Zant, T.Orlando, S.Watanabe and Strogatz, Phys. Rev. Lett. 74, 174
(1995); A.Ustinov, M.Cirillo and B.Malomed, Phys. Rev. B47, 8357 (1993);
Phys. Rev. B51, 3081 (1995).
 
11. S.Ryu, W.Yu and D.Stroud, Phys. Rev. E53, 2190 (1996); B.Kim, S.Kim
and
S.Lee, Phys. Rev. B51, 8462 (1995); J.Kim, W.Choe, S.Kim and H.Lee, Phys.
Rev. B49, 459 (1994).
 
12. R.A.Guyer and M.D.Miller, Phys. Rev. A17, 1205 (1978); {\it ibid}, A
17,
1774 (1978).
 
13. Y.Kivshar and B.Malomed, Rev. Mod. Phys. 61, 763 (1989).
 
14. S.Watanabe, H.Zant, S.Strogatz and T.Orlando, Physica D97, 429 (1996);
S.Watanabe, S.Strogatz, H.Zant and T.Orlando, Phys. Rev. Lett. 74, 379
(1995).
 
15. Zhigang Zheng, Bambi Hu and Gang Hu, CNS preprint No. 9710 (1997).
 
16. J.Currie, S.Trullinger, A.Bishop and J.Krumhansi, Phys. Rev. B 15,
5567
(1977).
 
17. M.Peyrard and M.Kruskal, Physica D 14, 88 (1984).
 
18. R.Boesch, C.R.Willis and M.El-Batanouny, Phys. Rev. B 40, 2284 (1989).
 
19. H.Risken, {\it The Fokker-Planck Equation, Methods of Solution and
Applications} (Springer-Verlag, Heidelberg, 1984).
 
20. Zhigang Zheng, Bambi Hu and Gang Hu (unpublished).
 
21. Zhigang Zheng and Gang Hu, Phys. Rev. E 52, 109 (1995).
 
22. Zhigang Zheng, Bambi Hu and Gang Hu, Phys. Rev. E 57, 1139 (1998).
 
\newpage 
 
\begin{center}
{\bf Caption of Tables}
 
\bigskip 
\end{center}
 
{\bf TABLE 1.} \ A comparison between numerical resonant steps and
theoretical predictions. All parameters are the same as those used in
Fig.1.
It can be clearly shown that the mean-field treatment predicts all
possible
resonances much better than previous predictions.
 
\bigskip\ 
 
\bigskip\ 
 
\begin{center}
{\bf Caption of figures}
 
\bigskip\ 
\end{center}
 
{\bf Fig.1.} The $v-F$ characteristics for $K=1.0$, $\gamma $$=0.1$, $N=8$
and $M=$ 1, 2, 3. Resonances can be observed on the step transitions. All
steps are labeled by using a pair of integers that describe the resonance
between the moving localized soliton and its radiated phonons. These
resonances are given by the formula (9). The agreement is quite good.
Several resonances are labeled on one step, because they are nearly
degenerated.
 
\medskip\ 
 
{\bf Fig.2.} The evolution of one particle in the chain for $K=1.0$,
$\gamma 
$$=0.1$ and different other parameters: (a) $N=16$, $M=1$, $F=0.05$; (b)
$%
N=8 $ , $M=3$, $F=0.02$; (c) $N=8$, $M=3$, $F=0.15$. Slip-stick motion can
be observed in (a) with high peak is the hopping from one well to another
well, low peaks is the influence of neighboring-particle hoppings, the
perturbation decays exponentially. (b) shows a $5:1$ resonance, while (c)
gives a high-order resonance.
 
\medskip\ 
 
{\bf Fig.3.} The $v-F$ relations for $N=13$, $M=8$, $\gamma $$=0.05$ and
$K=$%
2.0, 1.0, 0.25, 0.1. It is shown that resonant steps can be clearly
distinguished for larger coupling while the high- and low-velocity regimes
will connect for weak couplings, resonances will be smoothed and
disappear.
 
\medskip\ 
 
{\bf Fig.4.} The v-F relations for the approaching of $\delta $ to the
Golden mean $\delta _G=(\sqrt{5}-1)/2$ by using the Fibonacci sequence $%
\delta =\frac 58,\frac 8{13},\frac{13}{21},\frac{21}{34},...$ for $K=1.0$
and different dampings $\gamma =0.05,0.1,0.2.$ Different dampings do not
affect the step positions. Main steps for different cases agree well with
each other.
 
\medskip\ 
 
{\bf Fig.5.} The mean-field contraction factor $\beta $ varying with the
frustration $\delta $. The curves are symmetric about $\delta =1/2$. The
factor decreases when increasing the coupling K. Peaks are rational
resonances, which form a Farey tree.
 
\medskip\ 
 
{\bf Fig.6.} The maximum velocity $v_{max}$ varies with the frustration $%
\delta $ for the coupling strength $K=0.5$, 1.0 and 10.0. The effect of
contraction factor cannot be ignored for smaller $K$. Nonlinear regions
can
also be observed at the edges of the curves.
 
\medskip\ 
 
{\bf Fig.7.} Plot of the two functions $f(x)=\sin {}^2x$ and
$f(x)=x^2-\frac
\beta {4K}$ for $K=1.0$ and $\beta =$1.0, 0.5 and 0.0 (dotted lines from
below).
 
\medskip\ 
 
{\bf Fig.8.} The $v-F$ plot for $N=16$ and $M=1$, where $K=1.0$, $\gamma
$$%
=0.1$. Both low- and high-velocity regimes are shown. Resonances also
occur
in the high regime due to the whirling-mode induced parametric
instability.
Inset is the enlarged plot of the low-velocity regime. All resonances are
labeled by using the theoretical formula (9). Resonances $(m_1,1)$ that $%
m_1<6$ disappear due to the resonance prohibition criterion given by (13).
 
\medskip\ 
 
{\bf Fig.9.} A historic plot of the surface of section $\sin x_2-\sin x_1$
where the mean phase $\stackrel{-}{x}(t)=\frac 1N\sum_{j=1}^Nx_j(t)=0$
(mod $%
2\pi $) for $N=8$, $M=2$ on the $7:3$ step. $F=$0.112, 0.12, 0.13, 0.15,
0.16 and 0.165 for sections from (a) to (e), respectively. The torus
represents quasiperiodic motion. The size of the torus increases and
deforms
with increasing $F$ until it breaks into a new point that denotes a
periodic
motion, indicating the transition to another dynamical state.
 
\medskip\ 
 
{\bf Fig.10.} A cartoon of Poincare sections $\sin x_2-\sin x_1$ for
$N=8$, $%
M=3$ on the $3:2$ step, where $F$ varies from 0.162 to 0.213 for the
pictures from (a) to (l). It vividly demonstrates the bifurcation from
periodic motion to quasiperiodic and chaotic motions. When F increases,
the
torus becomes larger and twists, then forms a web and eventually becomes
chaotic.
 
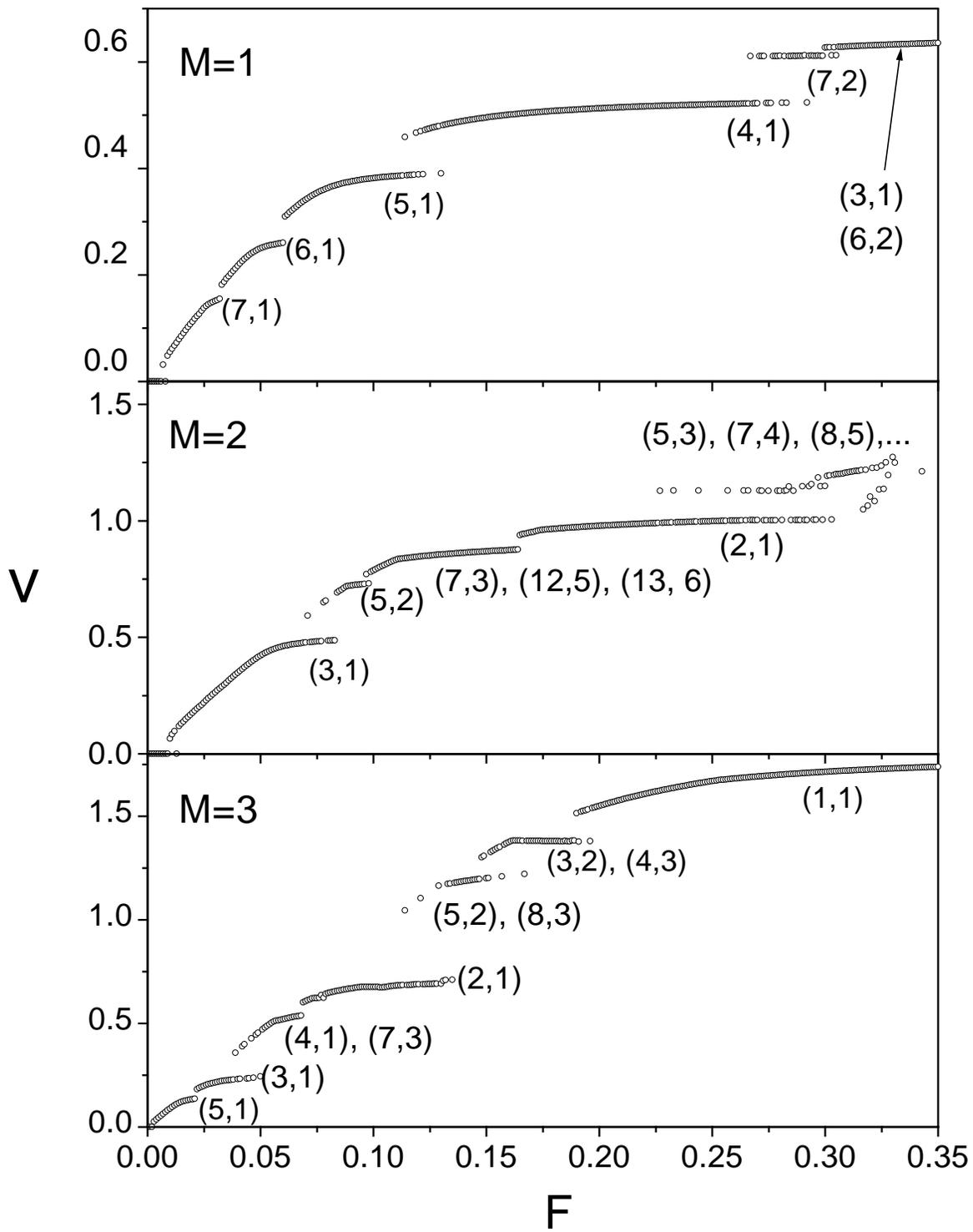
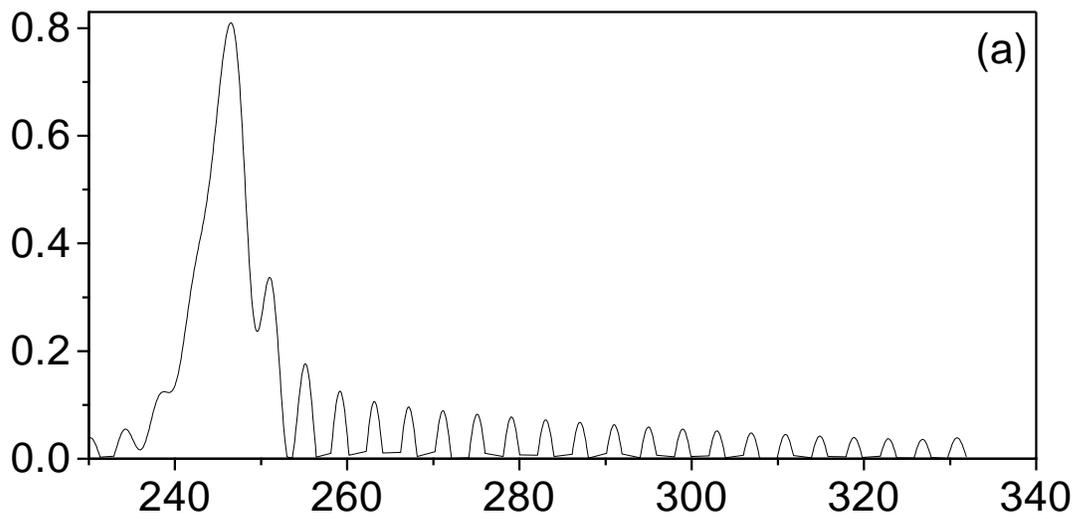
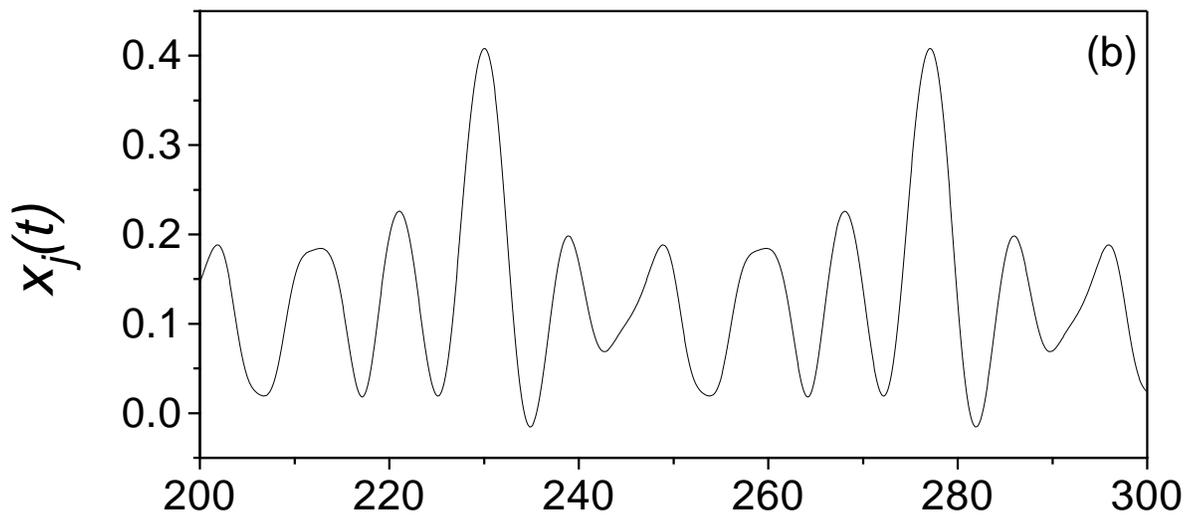
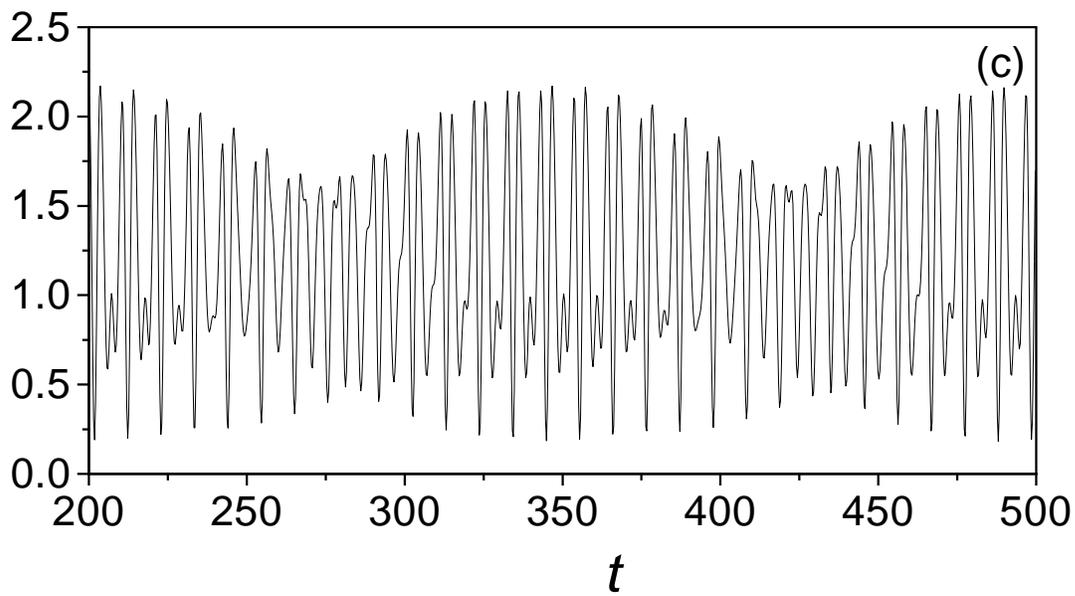
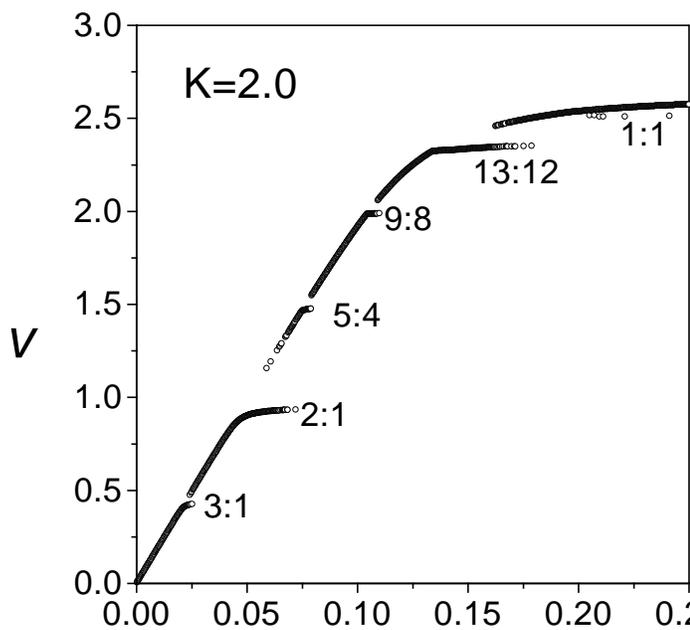
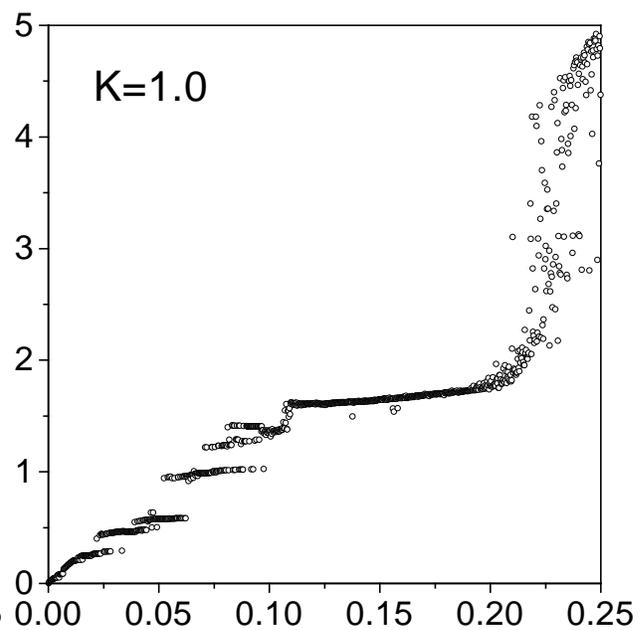
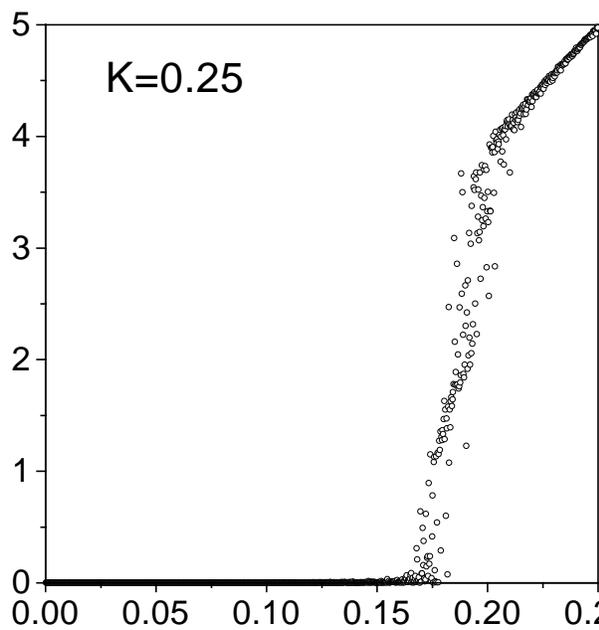
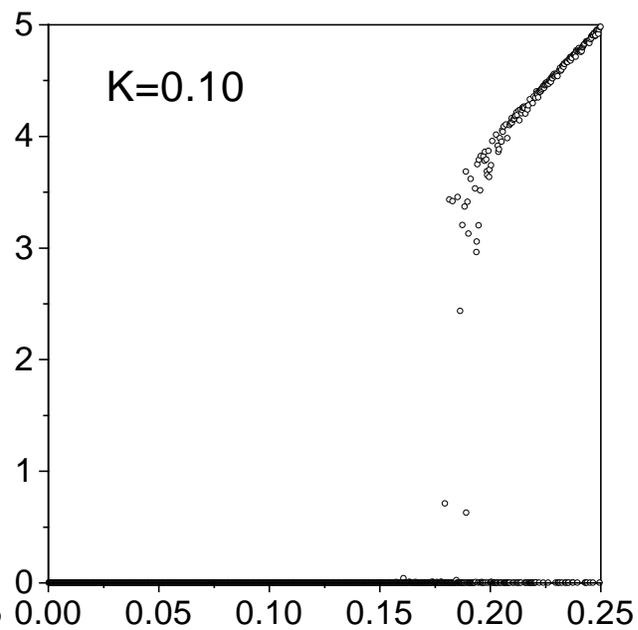
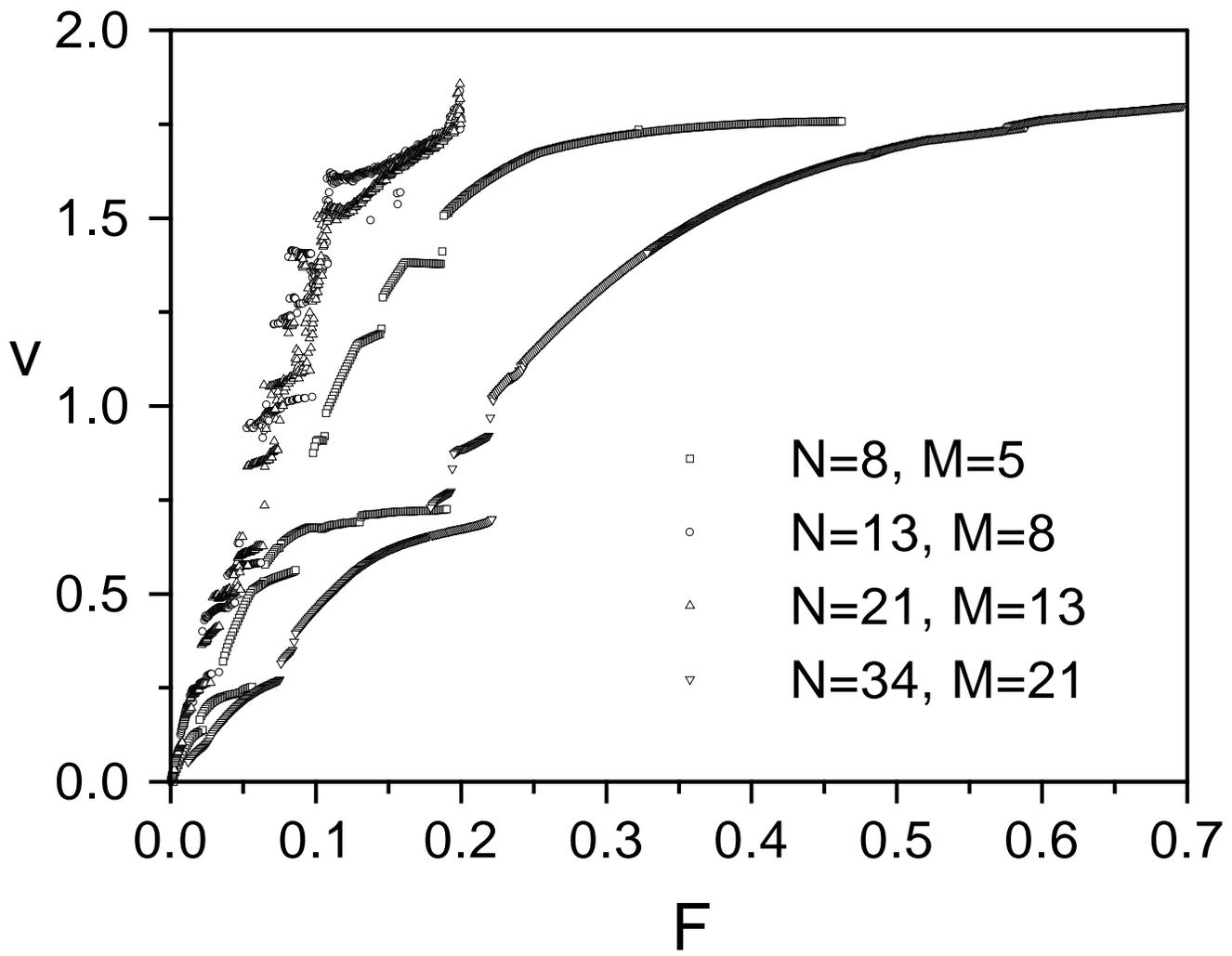
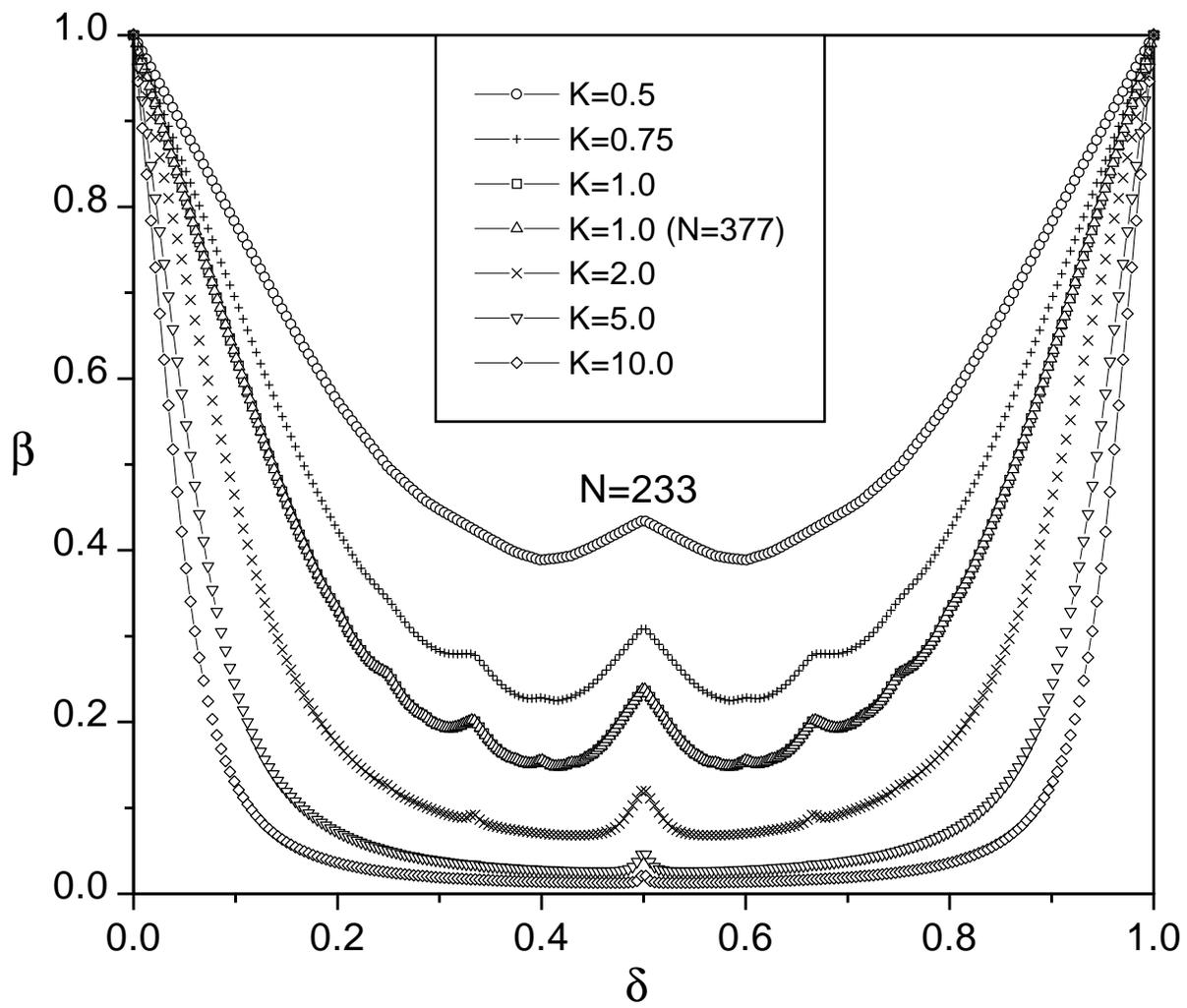
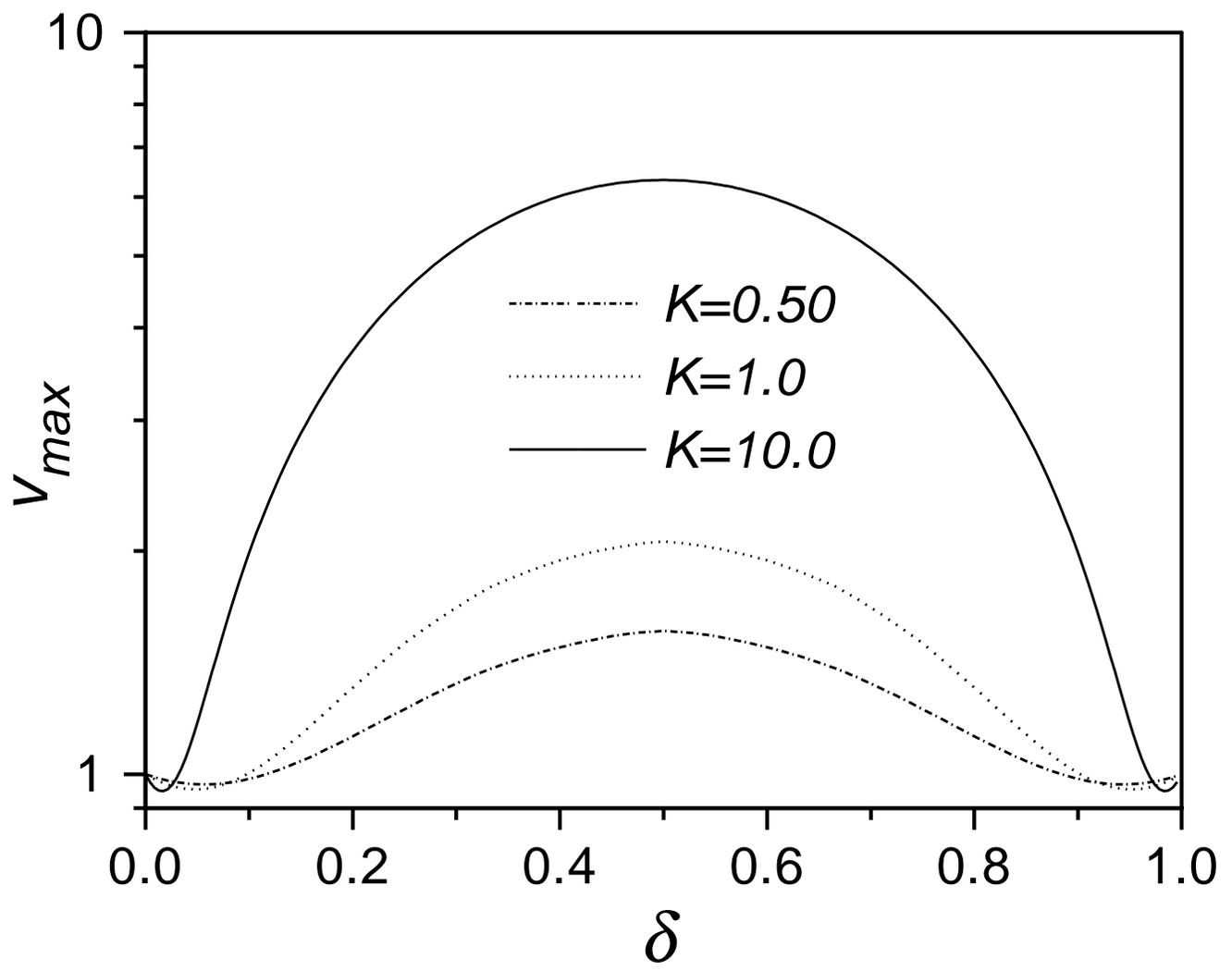
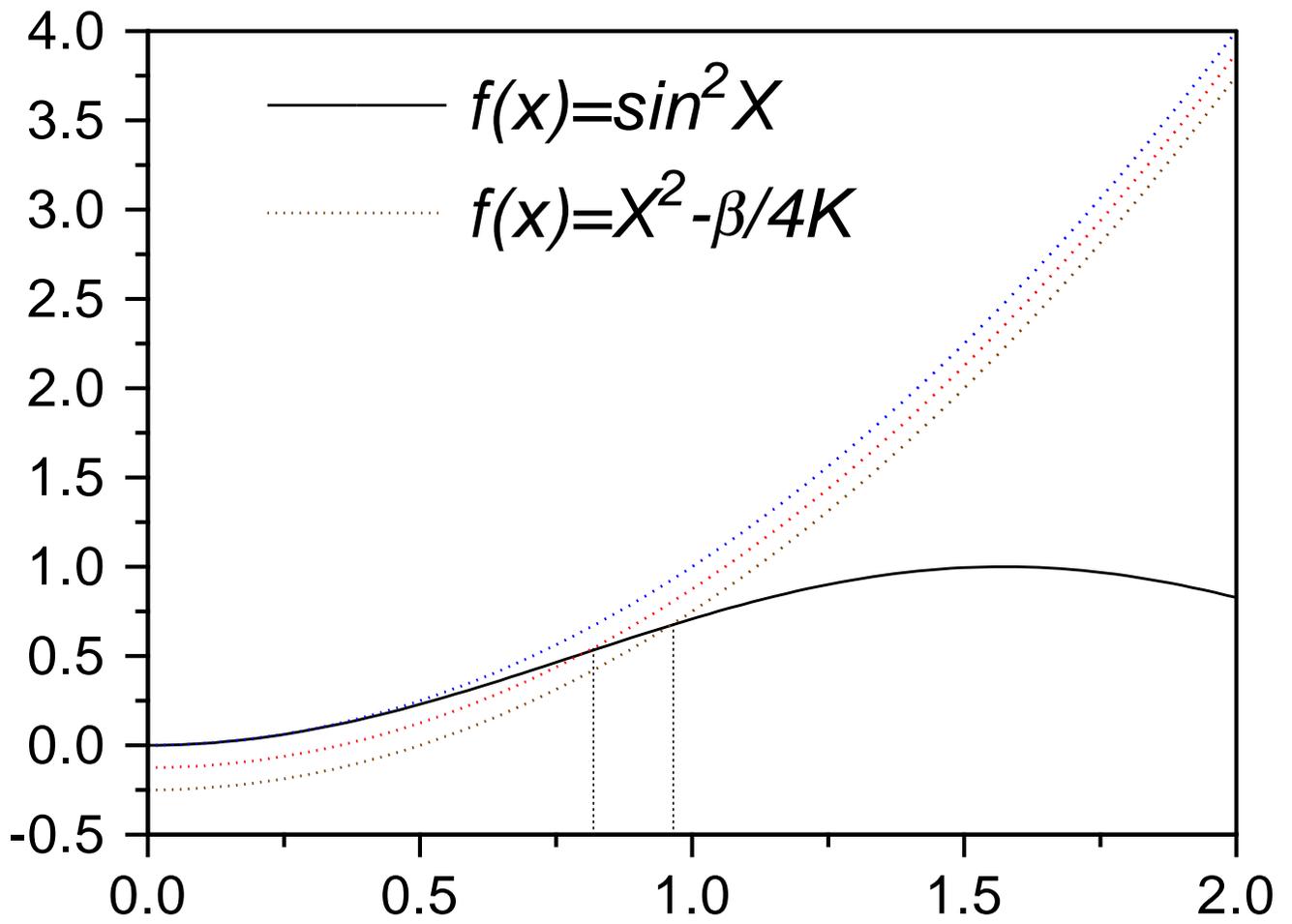
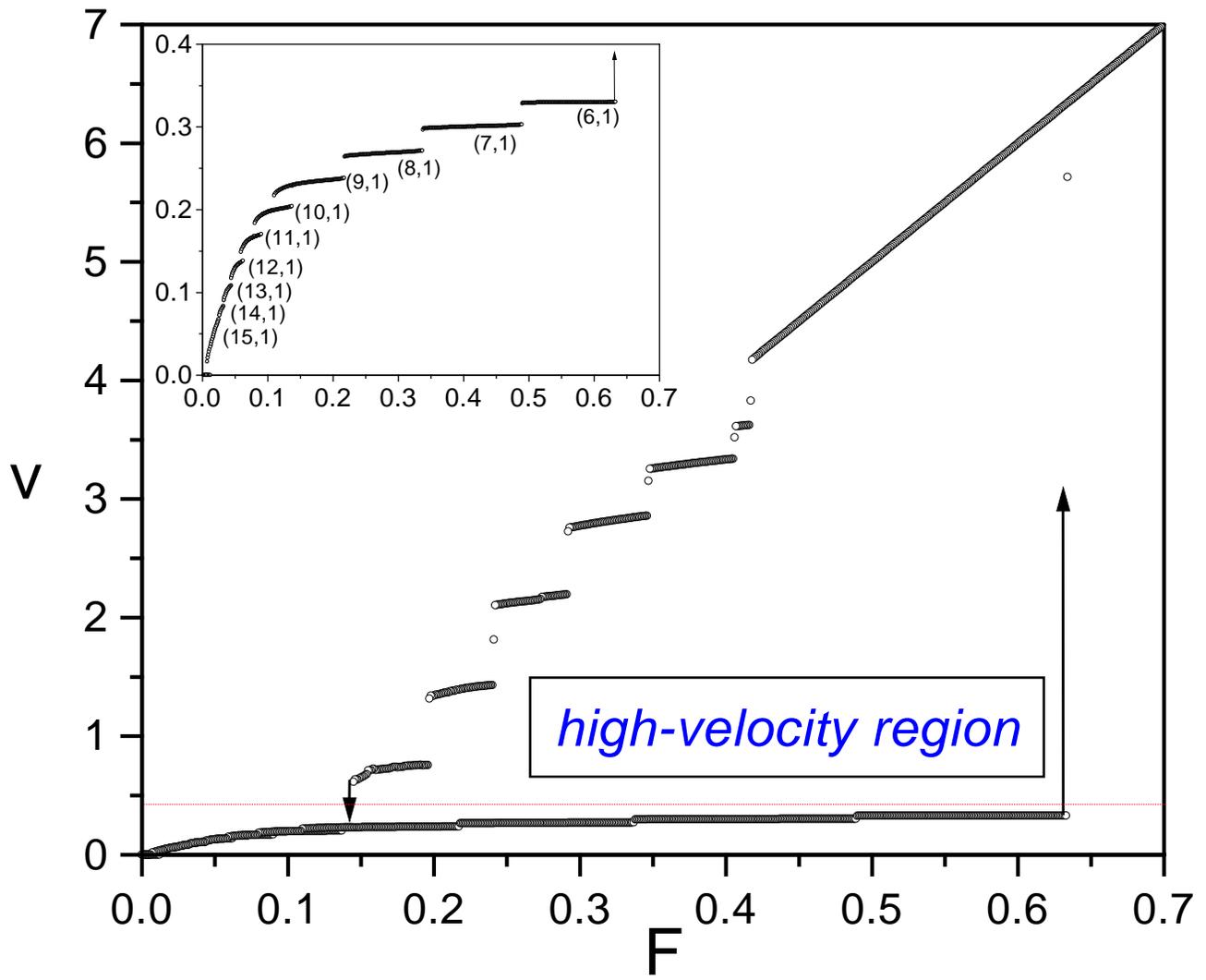
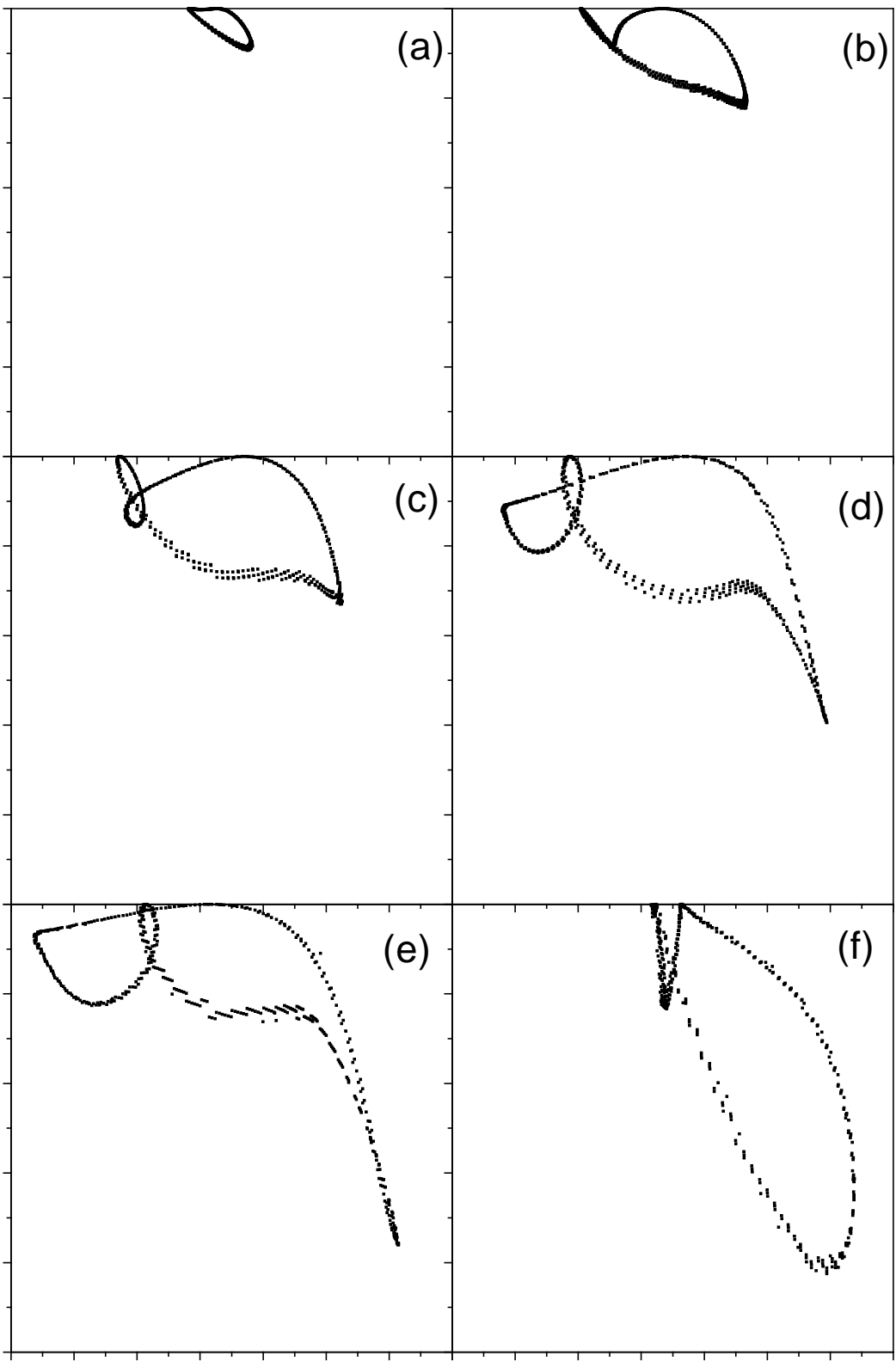
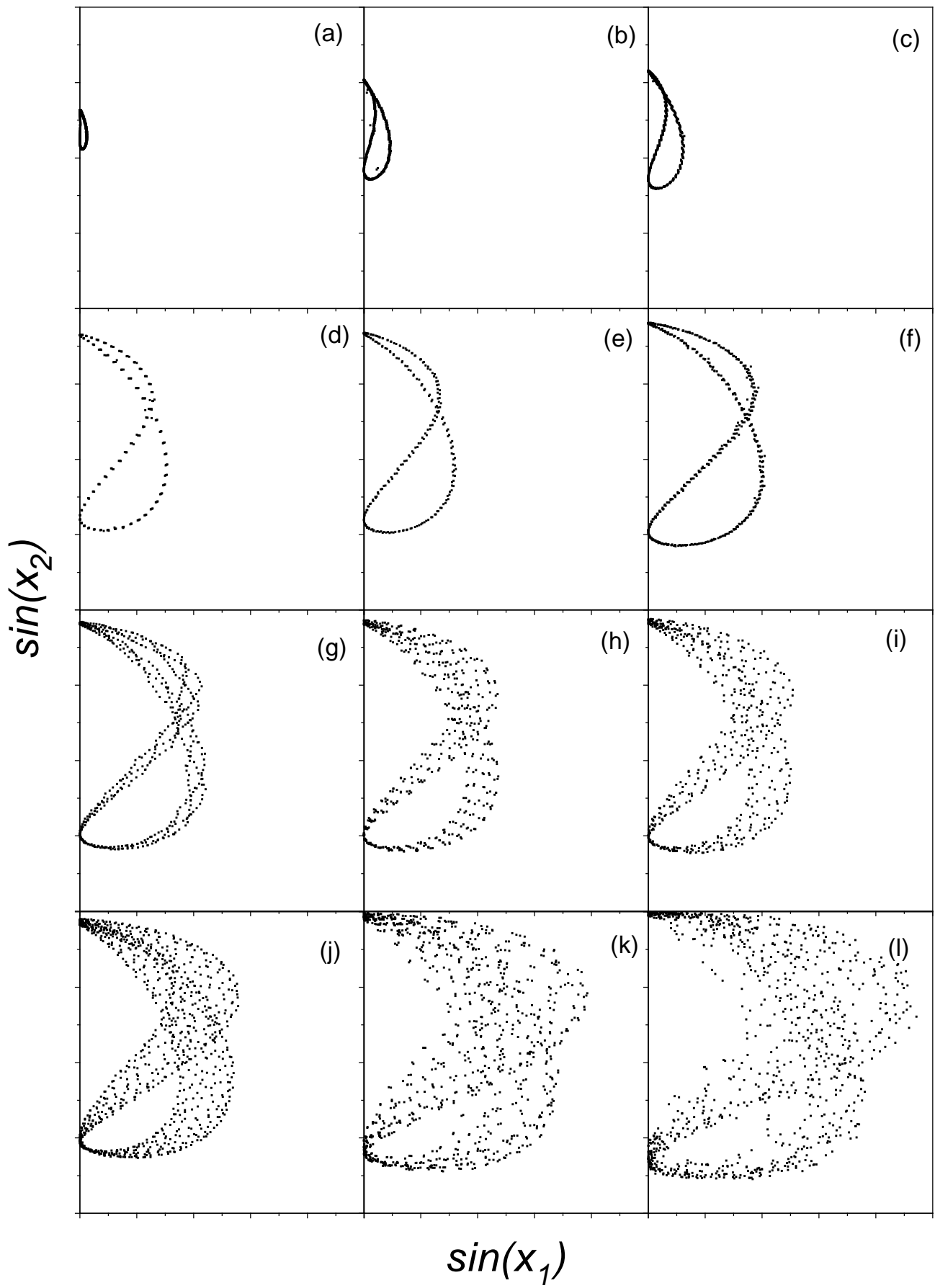
\end{document}